\documentclass[draft]{aipproc}

\layoutstyle{6x9}

\begin{document}

\title{GLAST's GBM Burst Trigger}

\author{D. Band}{address={Code 661, NASA/Goddard Space Flight Center, Greenbelt, MD  20771}}

\author{M. Briggs}{address={National Space Science and
Technology Center, Huntsville, AL 35805}}

\author{V. Connaughton}{address={National Space Science and
Technology Center, Huntsville, AL 35805}}

\author{M. Kippen}{address={NIS-2, Los Alamos National Laboratory, Los Alamos, NM 87545}}

\author{R. Preece}{address={National Space Science and
Technology Center, Huntsville, AL 35805}}

\begin{abstract}
The GLAST Burst Monitor (GBM) will detect and localize
bursts for the GLAST mission, and provide the spectral and
temporal context in the traditional 10~keV to 25~MeV band
for the high energy observations by the Large Area
Telescope (LAT). The GBM will use traditional rate triggers
in up to three energy bands, and on a variety of timescales
between 16~ms and 16~s.
\end{abstract}

\maketitle

\section{The Mission}
The Gamma-ray Large Area Space Telescope (GLAST) is the
next NASA general gamma-ray astrophysics mission, which is
scheduled to be launched into low Earth orbit in February,
2007, for 5--10 years of operation.  It will consist of two
instruments: the Large Area Telescope (LAT) and the GLAST
Burst Monitor (GBM).  A product of a NASA/DOE/international
collaboration, the LAT will be a pair conversion telescope
covering the $<$20~MeV to $>$300~GeV energy band. The LAT
will be $\sim$30 times more sensitive than {\it CGRO's}
EGRET.

The GBM will detect and localize bursts, and extend GLAST's
burst spectral sensitivity to the $<$10~keV to $>$25~MeV
band. Consisting of 12 NaI(Tl) (10--1000~keV) and 2 BGO
(0.15--25~MeV) detectors, the GBM will monitor $>$8~sr of
the sky, including the LAT's field-of-view (FOV).  Bursts
will be localized to $<15^\circ$ (1$\sigma$) by comparing
the rates in different detectors.

During most of the mission GLAST will survey the sky by
rocking $\sim35^\circ$ above and below the orbital plane
around the zenith direction once per orbit.  The first year
will be devoted to a sky survey while the instrument teams
calibrate their instruments.  During subsequent years guest
investigators may propose pointed observations, but
continued survey mode is anticipated because it will
usually be most efficient.

Both the GBM and the LAT will have burst triggers.  When
either instrument triggers, a notice with a preliminary
localization will be sent to the ground through TDRSS and
then disseminated by GCN.  Additional data will be sent
down through TDRSS for an improved rapid localization on
the ground.  ``Final'' positions will be calculated from
the full downlinked data. All positions will be
disseminated as GCN Notices, and additional information
(e.g., fluences and durations) will be sent out as GCN
Circulars.

Using its own and the GBM's observations, the LAT will
decide autonomously whether to slew to the burst location
for a 5~hour followup pointed observation. The threshold
will be higher for GBM-detected bursts outside the LAT's
FOV.

The GBM's NaI and BGO detectors will provide the number of
counts detected in 8~energy bands every 16~ms; the GBM will
trigger off these rates. Rate triggers test whether the
increase in the number of counts in an energy band $\Delta
E$ and time bin $\Delta t$ is statistically significant;
the expected number of non-burst photons in the $\Delta
E$--$\Delta t$ bin is estimated by accumulating counts
before the time bin being tested. Building on our
experience with the BATSE trigger, we are performing trade
studies to optimize the sensitivity of these triggers. Here
we present the results of our studies focusing on the
choice of $\Delta E$ and $\Delta t$.
\section{Choice of $\Delta$\hbox{\rm{\it t}}}
We consider two $\Delta t$ hierarchies---$\Delta t$ spaced
by factors of $\times$2 (e.g., 16 ms, 32 ms, 64 ms...) or
$\times$4 (e.g., 16 ms, 64 ms, 256 ms...)---and three time
bin registrations---non-overlapping bins (e.g., bins
separated by 1024 ms for $\Delta t$=1024 ms bins),
half-step bins (e.g., bins separated by 512 ms for $\Delta
t$=1024~ms bins), and all possible bins (e.g., bins every
16~ms). Varying $\Delta t$ can maximize the signal-to-noise
ratio, while more time registrations permit the bin to be
centered over the peak flux, also maximizing the
signal-to-noise ratio. To test these 6 triggers we applied
them to the 64~ms resolution lightcurves of the 25
brightest BATSE bursts; for each lightcurve we chose 10
starting times at random. Note that the GBM rates will have
16~ms resolution.

The most sensitive trigger would have $\Delta t$ spaced by
$\times$2 and every possible time bin.  The next most
sensitive trigger would have $\Delta t$ spaced by $\times$2
and bins spaced every half step.  These triggers would test
different numbers of bins---11264 bins vs. 3070 bins tested
every 16.384~s.

Besides the increased computational burden, the risk of a
false trigger increases as the number of bins tested
increases, but the false trigger probability is not
proportional to the number of bins because the bins are not
independent.  Our simulations indicate this is a $<$5\%
effect---for the same false trigger rate the trigger
threshold should be raised by a few percent as the number
of time bins tested increases.

\section{Choice of $\Delta E$}

Triggering on the counts accumulated in different $\Delta
E$ can tailor the detector sensitivity to hard or soft
bursts. The GBM will be able to trigger on more than one
$\Delta E$, and therefore we seek the set that will
maximize the sensitivity for both hard and soft bursts,
although hard bursts are a priority since their spectra are
more likely to extend into the LAT's energy band (of course
one must be careful of one's theoretical prejudices). For
the study of detector sensitivity to different types of
bursts and for comparisons between detectors, the
$F_T$-$E_p$ plane is useful\cite{1}, where $F_T$ is the
peak photon flux in a fiducial energy band (here
1--1000~keV) and $E_p$ is the energy of the peak of $E^2
N(E)\propto \nu f_\nu$. Burst spectra are also
characterized by asymptotic low and high energy power laws
with spectral indices $\alpha$ and $\beta$, respectively;
the dependence of $F_T$ on $\alpha$ and $\beta$ is not as
great as on $E_p$. For a given set of spectral indices the
detector sensitivity (the threshold value of $F_T$ at a
give $E_p$) is a curve in the $F_T$-$E_p$ plane.

To calculate these sensitivity curves we need both the
number of counts a detector will detect in the nominal
$\Delta E$ band for a given burst spectrum and the number
of background counts in this $\Delta E$.  A code has been
developed to calculate these numbers for each GBM detector
for a burst in any direction relative to the spacecraft.
Currently the code uses response matrices for the flux
directly incident on the detectors (without scattering off
the spacecraft or the Earth, but with obscuration by other
parts of the observatory), and a model of the background on
orbit. We used this code to calculate the sensitivity along
the normal to the LAT for $\Delta t$=1.024 s assuming at
least two detectors trigger at $\sigma_0>5.5$.

We calculated the sensitivity curves for a variety of
$\Delta E$.  The extremes of our spectral index sets were
$\alpha$=0, $\beta=-2$ and $\alpha=-1$, $\beta=-25$. The
first set is similar to the spectra sometimes observed
early in a burst; its high energy tail would be more easily
detected by the LAT. The second set is a spectrum without a
high energy tail.

\begin{figure}
  \includegraphics[height=.25\textheight]{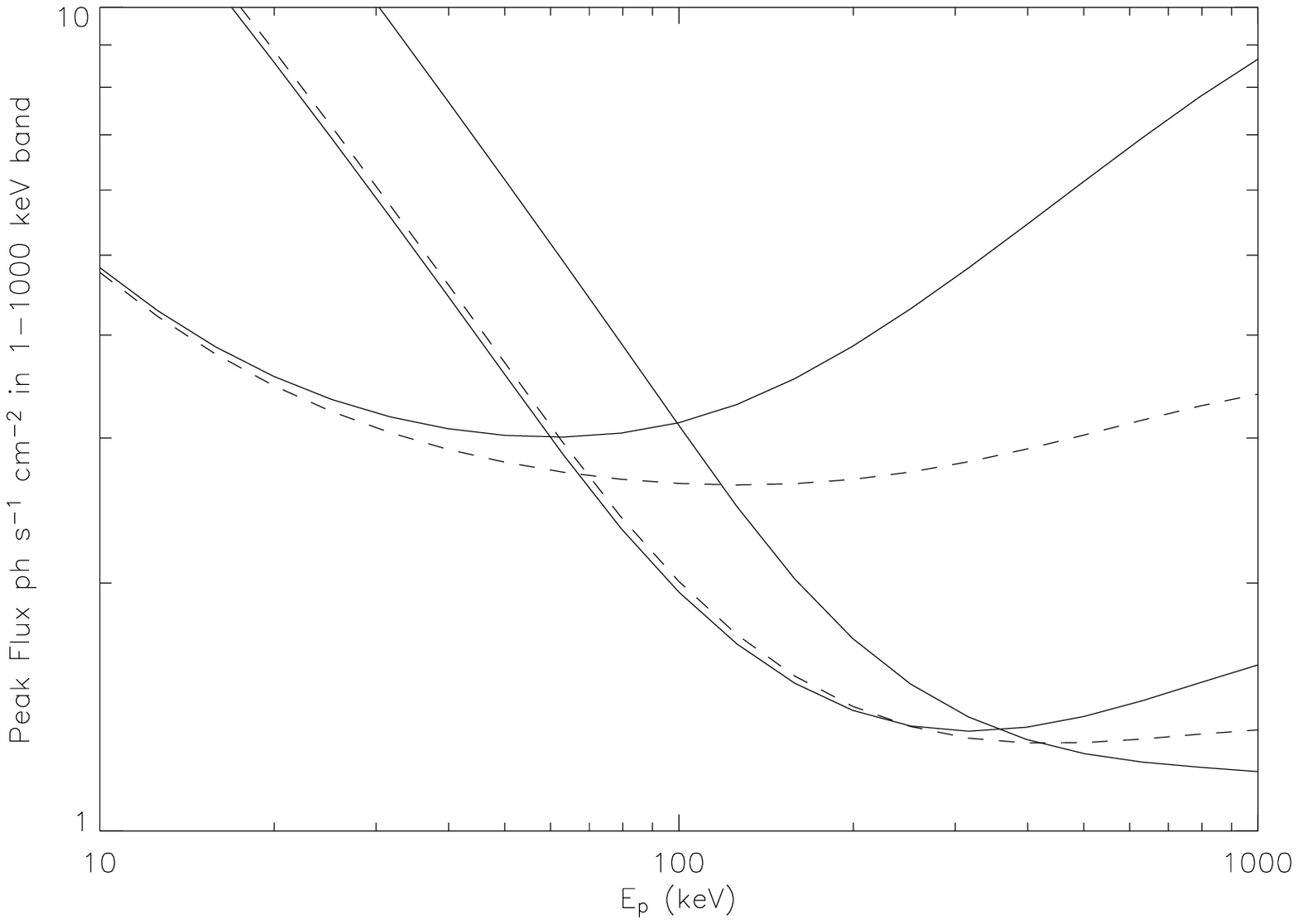}
  \includegraphics[height=.25\textheight]{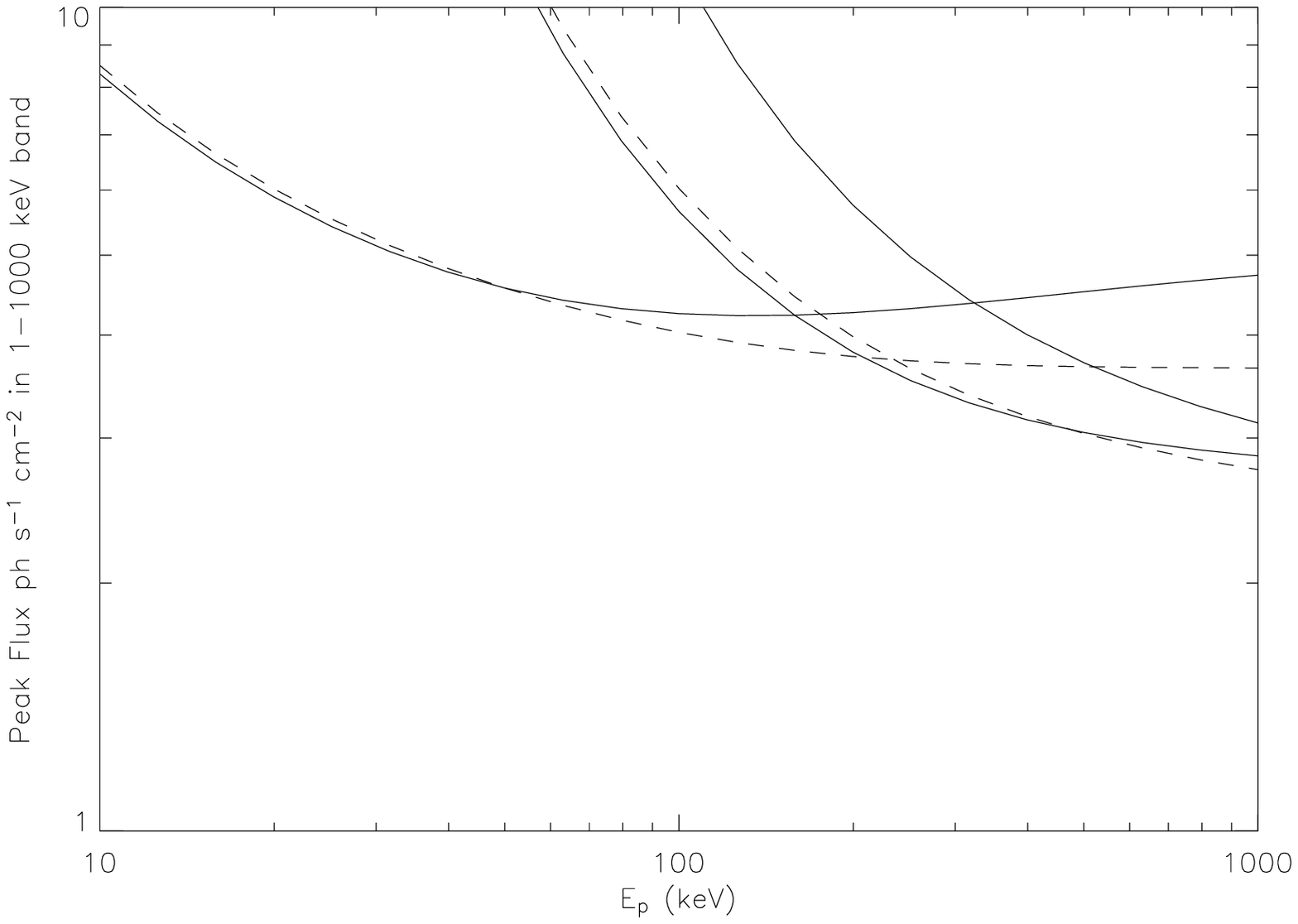}
  \caption{Sensitivity for two sets of $\Delta E$
for $\alpha$=0, $\beta=-2$ (left plot) and $\alpha=-1$,
$\beta=-25$ (right plot).  The solid curves are for (left
to right within plot) $\Delta E$=5--100, 50--300 and
100--1000~keV and the dashed for $\Delta E$=5--1000 and
50--1000~keV.  Lower curves are more sensitive.}
\end{figure}

Figure~1 shows the sensitivity for two sets of $\Delta E$.
To compare the GBM and BATSE burst distributions we would
like to include $\Delta E$=50--300~keV, which was BATSE's
primary trigger band. As can be seen, $\Delta E$ with a low
energy cutoff of $\sim$50~keV is optimal for high energy
sensitivity because it does not include the large low
energy background. Conversely, $\Delta E$ should always
extend to the highest energy possible because of the low
background at high energy.

We find that triggering a single BGO detector with
$\sigma_0=8$ increases the sensitivity for $E_p>1$~MeV for
$\alpha$=0, $\beta=-2$.  There is no increase in
sensitivity for $\alpha = -1$.

Figure~2 compares the $\Delta t$=1~s sensitivity for the
GBM (solid) and BATSE (dot-dashed) with the burst intensity
(dashed) that when the spectrum is extrapolated to the LAT
energy band would result in 25~detected photons per second.
The burst is assumed to be on the LAT normal, $\alpha=-1$,
$\beta=-2$, and the GBM triggers on $\Delta E$=5--100 and
50--300~keV.

\begin{figure}
  \includegraphics[height=.25\textheight]{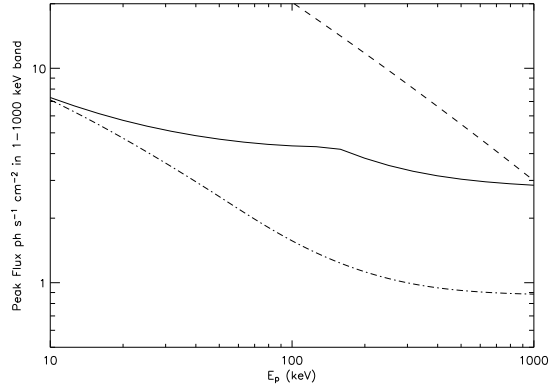}
  \caption{Comparison of GBM (solid curve) and BATSE (dot-dashed)
  sensitivities, and the flux necessary for a LAT detection
  of 25~photons (dashed).}
\end{figure}

\section{Conclusions}
The $\Delta t$ calculations suggest that spacing $\Delta t$
by factors of $\times$2 (i.e., 16~ms, 32~ms, 64~ms...) and
staggering the bins by half a timestep (e.g., the 1024~ms
bins are accumulated every 512~ms) would be particularly
efficient given the number of time bins that would be
tested. Choosing two triggers with $\Delta E$ starting at
5~keV and 50~keV and extending to the detector's high
energy cutoff would provide good high and low energy
sensitivity. Using $\Delta E$=50--300~keV would reproduce
the BATSE trigger, but would reduce the $E_p>$500~keV
sensitivity for the hardest bursts (which are more likely
to have LAT flux); this can be mitigated by adding $\Delta
E$=100--1000~keV.

Future trade studies will focus on the methodology for
measuring the background. GLAST will slew more frequently
than {\it CGRO} did, and therefore the background rates in
GLAST's different detectors will probably vary more
rapidly.  We will need to suppress spurious triggers
resulting from variations in the background without a major
reduction in the GBM's sensitivity.

Ultimately the trigger design will be constrained by the
computational capabilities of the GBM's processor.

\end{document}